\documentclass[12pt]{article}
\begin{document}
\begin{center}
{\large\bf $\mathcal{C}^{2}$ Formulation of Euler Fluid }\\
\vskip 1cm
{ G.P.Pronko}\\

{\it Institute for High Energy Physics , Protvino, Moscow reg.,
Russia,\\
Departmendo de Fisica Teorica, Atomica y Optica, Universidad de
Validated, 47071 Validated, Spain}

\end{center}

\begin{abstract}
The Hamiltonian formalism for the continuous media  is constructed
using the representation of Euler variables in $\mathcal{C}^{2}\times
\infty$ phase space.
\end{abstract}

\section{Introduction}

In the theory of continuous media, such as fluid, gas or plasma there
exist two kinds of description usually refereed as Lagrange and
Euler.The first uses the trajectories of the particles which
constitute the media, while in Euler description the role of
dynamical variables is played by the velocity $\vec v(x,t)$ and
density $\rho(x,t)$. As it was discussed in \cite{Pronko} the
Lagrange approach which uses the coordinates of the particles which
constitute the media, is very convenient to introduce the interaction
between the particles and for the construction of Hamiltonian
formalism, which looks like usual field theory canonical formalism
and the only problem is to construct the $x$-dependent canonical
variables. This problem was solved in \cite{Pronko} and the outcome
of the approach suggested there was a mechanical system where
evolution is described by canonical Hamiltonian equations in the
$2d\times\infty$-dimensional phase space (for the $d$-dimensional
fluid). Euler variables could be expressed via the canonical
variables and algebra of its Poisson brackets has the following form
for any dimension of space:
\begin{eqnarray}\label{1}
&\{v_j(x),v_k(y)\}&=-\frac{1}{m \rho(x)}\left(\nabla_j v_k(x)-
\nabla_k v_j(x)\right)\delta(\vec x-\vec y)\nonumber\\
&\{v_j(x),\rho(y)\}&=\frac{1}{m}\nabla_j\delta(\vec x-\vec
y)\nonumber\\
&\{\rho(x),\rho(y)\}&=0,
\end{eqnarray}
The parameter $m$ here is the mass of the particles which constitute
the fluid. Usually it does not appear in Euler description but here
it was inherited from Lagrange description, we put it equal to unity.
The relation of Lagrange and Euler description in fluid dynamics was
discussed in many text books and articles (see \cite{Salmon}, there
could be found numerous references to the earlier investigations,
\cite{Jackiw1},\cite{Arnold},\cite
{Lanczos}\cite{Pronko},\cite{Lamb},\cite{Mars}) and we will not
discuss it here. The typical Hamiltonian of ideal fluid could be
expressed in terms of Euler variables and it has the following form:
\begin{equation}\label{2}
H=\int d^3 x[\frac{1}{2}m \rho(x)\vec v(x)^2+V(\rho(x))],
\end{equation}
where the function $V(\rho(x_i)$ describes the "potential" energy of
the fluid and could be chosen phenomenologically \cite{Zakharov}:
\begin{equation}\label{3}
V(\rho (x))=\frac{\kappa}{2\rho_0} (\delta \rho (x))^2 +\lambda
(\nabla \rho (x))^2+... \quad,
\end{equation}
where $\delta \rho (x)$ is the deviation of the density from its
homogeneous distribution $\rho_{as}$ at infinity:
\begin{equation}\label{4}
\delta \rho (x)=\rho (x)-\rho_{as}.
\end{equation}
The first term in (\ref{3}) is responsible for the  sound wave in the
fluid ($\kappa$ is the velocity of sound), the second term in
(\ref{3}) describes the  dispersion of the sound waves. The equations
of motion for the variables $\vec v(x,t),\rho (x,t)$ could be written
in a canonical form:
\begin{eqnarray}\label{5}
\dot{\vec v}(x,t)=\{H,\vec v(x,t)\},\nonumber\\
\dot{\rho}(x,t)=\{H,\rho(x,t)\}.
\end{eqnarray}
The Poisson brackets to used in (\ref{5}) are given by (\ref{1}). The
situation we have described looks quite standard for a mechanical
system , but the point is that the variables $\vec v(x,t),\rho(x,t)$
do not belong to the phase space because  the Poisson brackets
(\ref{1}) are degenerate and we can not consider it as usual
coordinates of the phase space . This kind of Poisson algebra could
be treated by Kirillov-Konstant approach \cite{KK}. The center of the
algebra (\ref{1}) depends of the dimension of configuration space.
For example, in 2-dimensional case the center is formed by:
\begin{equation}\label{6}
I_n=\int d^2 x \rho(x)^{1-n}(\partial_1 v_2(x)-\partial_2 v_1(x))^n,
\end{equation}
see \cite{Arnold}, \cite{Jackiw1} for discussion. The 3-dimensional
case we shall consider in details below. Here one of the  Casimirs is
the "helicity" functional
\begin{equation}\label{7}
Q=\int d^3 x \,\epsilon_{jkl}\, v_j(x)\nabla_k v_l(x).
\end{equation}
The other Casimir is the total number of particles $N$(valid for any
dimension, for $d=2,N=I_0$):
\begin{equation}\label{8}
N=\int d^3 x \rho(x)
\end{equation}

\section{ $C^2$ Hydrodynamics}

The 3-dimensional case, which is very important for applications was
considered by many authors starting from IXX century. It is hardly
possible to give an exhaustive list of references. Recently it was
discussed in  \cite{Jackiw1} (see also \cite{Salmon} for earlier
references) where it was suggested to build Hamiltonian formalism for
3-dimensional Euler fluid using Clebsh parameterization \cite{Clebsh}
for the velocity :
\begin{equation}\label{4.1}
\vec v(x,t)=\vec \partial \alpha (x,t)+\beta \vec \partial
\gamma(x,t)
\end{equation}
where the new functions $\alpha(x,t),\beta(x,t),\gamma(x,t)$ together
with density $\rho(x,t)$ are used for the construction of the
coordinates of the phase space. What we are going to suggest here is
an alternative approach, which has certain advantages.

Let us consider a mechanical system which is described by a pair of
complex coordinates which belong to $C^2\times\infty$:\quad
$u_{\alpha} (x,t),\quad \bar u_{\alpha} (x,t)$, where $\alpha=1,2$.
The Lagrangian for this system we shall take in the following form:
\begin{eqnarray}\label{4.2}
L&=&\int d^3 x\{\frac{i m}{2}(\bar u (x,t) \dot u (x,t)-\dot{\bar
u} (x,t)u (x,t)) \nonumber \\
&+&m\frac{(\bar u (x,t)\vec \partial u (x,t)-\vec\partial\bar u
(x,t)u (x,t))^2}{8 \bar u (x,t)u (x,t)} -V(\bar u (x,t)u (x,t))\},
\end{eqnarray}
where we assume the summation over indexes. The canonical momenta,
corresponding to the variables $u_{\alpha} (x),\quad \bar u_{\alpha}
(x)$ are given by equations
\begin{eqnarray}\label{4.3}
p^{u}_{\alpha} (x)&=&\frac{i m}{2}\bar u_{\alpha} (x), \nonumber \\
p^{\bar u}_{\alpha} (x)&=&-\frac{i m}{2}u_{\alpha} (x).
\end{eqnarray}
As it expected for the Lagrangian which is a  linear function of
velocities, the equations (\ref{4.3}) define the constraints on the
canonical variables:
\begin{eqnarray}\label{4.4}
\lambda_{\alpha}^1 (x)=p^{u}_{\alpha} (x)-\frac{i m}{2}\bar
u (x)_{\alpha}\sim 0, \nonumber \\
\lambda_{\alpha}^2 (x)=p^{\bar u}_{\alpha} (x)+\frac{i m}{2}u
(x)_{\alpha}\sim 0.
\end{eqnarray}
The Poisson brackets of the constraints are non-degenerate
\begin{equation}\label{4.5}
\{\lambda_{\alpha}^1 (x),\lambda_{\beta}^2(y_i)\}=i m\delta_{\alpha
\beta}\delta(\vec x-\vec y)
\end{equation}
and we can use these constraints to eliminate canonical momenta
$p^{u}_{\alpha} (x),p^{\bar u}_{\alpha} (x)$ using Dirac procedure
\cite{Dirac}. The resulting Poisson (Dirac) brackets for the rest of
coordinates of the phase space $\tilde{\Gamma}$ are:
\begin{eqnarray}\label{4.6}
\{u_{\alpha} (x), \bar u_{\beta}(y_i)\}&=&\frac{i}{m}\delta_{\alpha
\beta}\delta(\vec x-\vec y),\nonumber\\
\{u_{\alpha} (x),  u_{\beta}(y_i)\}&=&0,\nonumber\\
\{\bar u_{\alpha} (x), \bar u_{\beta}(y_i)\}&=&0.
\end{eqnarray}
The Hamiltonian, corresponding to the Lagrangian (\ref{4.2}) has the
following form
\begin{equation}\label{4.7}
H=\int d^3 x[-m\frac{(\bar u (x)\vec \partial u (x)-\vec\partial\bar
u (x)u (x))^2}{8 \bar u (x)u (x)} +V(\bar u (x)u (x))]
\end{equation}

Now we shall explain why we consider this system. Let us form the
following objects:
\begin{eqnarray}\label{4.8}
\vec v(x)&=&\frac{1}{2i}(\bar u (x)\vec \partial
u (x)-\vec\partial\bar u (x)u (x)),\nonumber \\
\rho (x)&=&\bar u (x)u (x).
\end{eqnarray}
The notations we have used here are not accidental. The point is that
if we shall calculate the Poisson brackets for (\ref{4.8}), using
(\ref{4.6}) the result will exactly coincide with (\ref{1}). The
Hamiltonian $H$, given by (\ref{4.7}), being expressed via $\vec
v(x)$ and ${\rho (x)}$ will take the following form:
\begin{equation}\label{4.9}
H=\int d^3 x[\frac{1}{2}m \rho (x)\vec v(x)^2+V(\rho (x))],
\end{equation}
which also coincides with  Hamiltonian given by (\ref{2}).

The equations of motion for the variable $u_{\alpha} (x), \bar
u_{\alpha} (x)$ have the usual form:
\begin{eqnarray}\label{4.10}
\dot u_{\alpha} (x)=\{H,u_{\alpha} (x)\}\nonumber\\
\dot {\bar u}_{\alpha} (x)=\{H,\bar u_{\alpha} (x)\}
\end{eqnarray}
Apparently, the correct equations of motion , including the
continuity equation for variables $\vec v(x)$ and ${\rho (x)}$ follow
from (\ref{4.10}).

As was mentioned above, the description of the fluid in terms of
$u_{\alpha} (x),\quad \bar u_{\alpha} (x)$ is rather similar to the
description which uses  Clebsh parameterization. Indeed, these
variables could be presented in the following form:
\begin{equation}\label{4.11}
u_{\alpha} (x)=\displaystyle\sqrt{\rho (x)}e^{i \phi
(x)/2}\left(\begin{array}{c}
e^{-i\psi (x)/2}cos\frac{\alpha (x)}{2}\\
e^{i\psi (x)/2}sin\frac{\alpha (x)}{2}
\end{array}\right)\
\end{equation}
from where we obtain the representation for the velocity through
angles $\phi (x), \psi (x)$ and $\alpha (x)$
\begin{equation}\label{4.12}
\vec v(x)=\frac{1}{2}(\vec \partial \phi (x)-\vec \partial \psi
(x)cos\alpha (x))
\end{equation}
These equation defines the velocity, if Clebsh parameters are known.
Also, as is well-known  (see e.g.\cite{Lamb},\cite{Jackiw1}) any
differentiable vector field $\vec v(x)$ has the local representation
(\ref{4.12})\quad. In other words, knowing $\vec v(x)$, we can
construct Clebsh parameters $\alpha (x),\phi (x),\psi (x)$ with some
ambiguity. This ambiguity arises as a set of integration constants .
In  our construction this ambiguity could be understand as follows.
The Lagrangian function (\ref{4.2})  we consider is invariant with
respect to the symmetry group $U(2)$ which acts as follows:
\begin{eqnarray}\label{4.13}
u_{\alpha} (x)\rightarrow \tilde{u}_{\alpha} (x)=T_{\alpha
\beta}u_{\beta} (x), \quad T^{+}T=1
\end{eqnarray}
and according to Noether's theorem the integrals of motion, which is
the generators of these transformations are :
\begin{equation}\label{4.14}
t^{0}=\int d^{3}x\frac{1}{2}\bar{u} (x)u (x),\quad t^{a}=\int d^{3}x
\bar{u} (x)\frac{\sigma^{a}}{2} u (x)
\end{equation}
The transformations (\ref{4.14}) change the Clebsh variables, but
does not affect the Euler's variables. So, in particular, the
constant shift of the angle $\phi (x)$ is generated by ex-Casimir
$N$, which in $\tilde{\Gamma}$ has lost its status, the generator
$t^{3}$ shifts the angle $\psi (x)$ , the other two generators mix
the angles $\psi (x)$ and $\alpha (x)$ . So the system described by
variables $(u_{\alpha} (x), \bar{u}_{\alpha} (x))$ is a Hamiltonian
system with symmetry  and we can reduce its phase space by procedure
given by Souriau \cite{Sour} and Marsden and Weinstain \cite{MarsW} .
The reduced phase is the space, where "live" almost all the Euler
variables. The latter means that  the  procedure of reduction implies
fixing the integrals of motion, in particular $t^{o}=\frac{N}{2}$
does not anymore belongs to the set of variables.

The only problem we have now is   the "helicity" functional, which
still the is the Casimir and the reduction did not remove it. For
finite dimensional systems the existence of Casimir implies the
degeneracy of Poisson brackets. It could be easily seen from the
following consideration. By definition the Casimir $C$ should has a
vanishing brackets with all variable
\begin{equation}\label{4.15}
\{p_{k},C\}=0, \quad \{q_{k},C\}=0
\end{equation}
where $p_{k},q_{k}$ are all set of coordinates of the phase space. If
the Poisson brackets are non-degenerate, the equations (\ref{4.15})
mean that $C$ is a constant. The situation for the infinite
dimensional system is different because of existence of so called the
functionals with zero variation . Consider for example an infinite
dimensional system , which is described by the set of canonical
variables $p(x), q(x)$ where $x\in R$ . The Poisson brackets are
non-degenerate:
\begin{equation}\label{4.16}
\{p(x),q(y)\}=\delta(x-y).
\end{equation}
In this case we can easily construct a nontrivial functional, which
will have vanishing Poisson brackets with all variables $p(x), q(x)$.
It has the following form:
\begin{equation}\label{4.17}
C=\int_{-\infty}^{\infty} dx
\frac{p'(x)q(x)-p(x)q'(x)}{p^{2}(x)+q^{2}(x)}
\end{equation}
and has a meaning of a winding number for the phase of the complex
variable $a(x)=p(x)+i q(x)$, i.e. $C$ is what physicists used to call
topological charge. Note that in order for $C$ to be the Casimir it
is not necessary to impose the condition on
$a(x)|_{x\rightarrow-\infty}=a(x)|_{x\rightarrow+\infty}$ and
compactify $R$. In this case $C$ will take an integer values and
indeed will be the winding number.

The "helicity" functional has the same origin as the functional $C$
in this example. In order to see it let us introduce a unit four
vector $F_k$ \cite{FaddeevNiemi2}:
\begin{equation}\label{4.18}
\frac{1}{\sqrt{\bar{u} (x)u (x)}}\left( \begin{array}{c} u_1 (x)\\u_2
(x)\end{array}\right)=\left(
\begin{array}{c} F_1 (x)+iF_2 (x)\\F_3 (x)+iF_4 (x)\end{array}\right)
\end{equation}
This four vector maps $S^{3}\rightarrow S^{3}$, provided we impose on
the variables $\bar{u}_{\alpha} (x), u_{\alpha} (x)$ the asymptotic
conditions : $u_{\alpha}(x)\rightarrow u^{0}_{\alpha}$, when $|\vec
x|\rightarrow \infty$ and compactify $R^{3}$. The helicity functional
$Q$ given by (\ref{7}) could be written in the following form:
\begin{equation}\label{4.19}
Q=\frac{1}{3}\int d^{3}x \epsilon_{abcd}\epsilon_{ijk}F_a\partial_i
F_b\partial_jF_c\partial_kF_d,
\end{equation}
which is the standard representation (up to normalization constant)
for the winding number of the map $S^{3}\rightarrow S^{3}$, so called
Hopf invariant. Here again we should note that even if we neglect the
asymptotic conditions on $u_{\alpha}(x)$ together with
compactification of $R^{3}$, $Q$ still will be invariant with respect
to local variations and therefore will have vanishing Poisson
brackets with $\bar{u}_{\alpha} (x), u_{\alpha} (x)$. So, the
conclusion of this arguments is that for infinite dimensional
mechanical systems the existence of Casimirs is not necessary implies
the degeneracy of Poisson brackets, provided these Casimirs are
related to the geometric properties of the phase space. The helicity
functional belongs to this class of "friendly" Casimirs.

The description we presented here is very convenient for different
kinds of applications and generalizations. First of all it seems
rather convenient for investigation of stability  problem, because
the second variation of Hamiltonian (\ref{4.7}) is terms of the
variables $\bar{u}_{\alpha} (x), u_{\alpha} (x)$ is quite compact and
simple. As the generalizations are concerned, one can consider the
"relativization" of the Lagrangian (\ref{4.2}), supersymmetric
extension of the variables $\bar{u}_{\alpha} (x), u_{\alpha} (x)$ .
Increasing the number of components we can consider the media with
internal degrees of freedom et cetera. Also this formulation is very
convenient for quantization of fluid.  We are planning to present
some of these subjects in future publications.

\section{ Acknowledgements.}

I would like to thank professors  A.V. Rasumov ,L. Nieto and J. Negro
for valuable discussions, professors M. Gadella, J. Negro and L.
Nieto for kind hospitality in University of Validated . The work was
supported in part by the Russian Science Foundation Grant
04-01-00352, by Program for Support of Leading Scientific Schools
Grant 1303.2003.2 and by the Spanish MEC (grant SAB2004-0169)

\end{document}